\documentstyle{lamuphys}
\makeatletter
\let\chapter\hid@chapter
\makeatother

\begin{document}

\pagestyle{empty}

\addtolength{\baselineskip}{.2cm}

\begin{flushright}
UR-1528\\
ER/40685/916\\
May 1998
\end{flushright}

\vspace*{.2in}

\begin{center}
{\bf Standard-Model-like Chiral Spectra 
and the Origin of the Families}

\vspace*{.4in}
{\bf Otto C. W. Kong}

\vspace*{.4in}
{\it Department of Physics and Astronomy,\\
University of Rochester, Rochester NY 14627-0171.}

\vspace*{.5in}
{Abstract}
\end{center}

\noindent
We outline our approach to understand the family structure through the idea 
of a SM-like chiral fermion spectrum, one that is derivable from anomaly 
cancellation conditions in  the same way as the one family SM, under an 
extended symmetry, which after breaking to the SM symmetry yields the three 
families as the residual chiral content. An example of a relatively simple 
scalar sector which gives an acceptable symmetry breaking pattern and 
naturally hierarchical quark mass matrices is also discussed for a 
successful $SU(4)_A\otimes SU(3)_C\otimes SU(2)_L\otimes U(1)_X$ model.
This is a summary of a seminar given at the 37th IUKT winter school, for 
the proceedings.

\clearpage

\addtolength{\baselineskip}{-.2cm}

\pagenumbering{arabic}
\title{Standard-Model-like Chiral Spectra 
and the Origin of the Families}

\author{Otto C. W. Kong}

\institute{Department of Physics and Astronomy,
University of Rochester, Rochester NY 14627.}

\maketitle
%

Each family of the standard model (SM) fermions composes of
a chiral spectrum tightly bounded by gauge anomaly cancellaton constraints.
However, the three family structure and the hierarchy among the masses
is a major puzzle.

The  fermions in one SM family  can be uniquely derived by assuming
one multiplet transforming nontrivially under each component gauge group
and requiring the minimal chiral spectrum canceling all the anomalies.
We seek to understand the family structure through the idea of a SM-like 
chiral fermion spectrum, one with the same feature as the one family SM
under an extended symmetry, which after breaking to the SM symmetry
yields naturally the three families as the residual chiral content. The natural 
choice of the gauge group is $SU(N)\otimes SU(3)\otimes SU(2)\otimes U(1)$.
For instance, start with ${\bf (4,3,2,1)}$  for $N=4$, 
the strategy leads to the spectrum
\begin{center}
${\bf (4,3,2,1)}, {\bf (\bar{4},\bar{3},1,x)}, {\bf (\bar{4},1,2,y)}, 
{\bf (\bar{4},1,1,z)},$  \\
${\bf (1,\bar{3},2,a)}, {\bf (1,\bar{3},1,b)}, {\bf (1,\bar{3},1,c)},
{\bf (1,1,2,k)}, {\bf (1,1,1,s)}\; .$
\end{center}
Solution for the $U(1)$ charges canceling all anomalies exists but fails to 
give the correct SM embedding. However, analysis of the potentially successful 
embeddings of the three families suggests that
$SU(3)_C\otimes SU(2)_L\otimes U(1)_Y \; \subset\;
 SU(4)_A\otimes SU(3)_C\otimes SU(2)_L\otimes U(1)_X$
works when the above spectrum is augmented with  an anomaly-free
$SU(4)_A$ multiplet charged under $U(1)_X$. The resulted models have nontrivial
$U(1)_Y$ embeddings. Similar constructions with some other $N$ values are also 
obtained.

These SM-like chiral models have interesting phenomenological predictions.
In the case of a specific model with $N=4$, we also constructed a Higgs
sector giving  rise to a natural mass hierarchy
$m_t  ,   m_b  >  m_c  >   m_s >  m_d , m_u$.
The scalars multiplets are $\phi_0 = {\bf (\bar{4},1,1,9)}$ and
 $\phi_a$ ($a=1$ or $2$), in ${\bf (\bar{4},1,1,-3)}$, together with
$SU(2)_L$ doublets $\Phi = {\bf (15,1,2,-6)}$. A
$C_{ab} \phi_{ai} \phi_b^{\dag  j} \Phi^k_j \Phi^{\dag i}_k$ mass term
with natural  VEVs for the $\phi_a$'s decouples twelve of the fifteen doublets
from the EW-scale. There remain two EW Higgs doublets and an extra
doublet of singly- and doubly-charged scalars. FCNC constraints can be 
easily satisfied and the quark mass hierarchy resulted.

%
%


\begin{thebibliography}
%
%
\bibitem{}{k1}{}
Kong, \, O.C.W. (1996), 
 Mod. Phys. Lett. A 11, 2547.
\bibitem{}{k2}{}
Kong, \, O.C.W. (1997), 
 Phys. Rev. D 55, 383.
\end{thebibliography}
\end{document}